\documentclass[a4paper,fleqn]{cas-sc}

\usepackage{comment}
\usepackage[square,numbers, sort&compress]{natbib}
\usepackage{hyperref}

\usepackage{xurl}
\hypersetup{breaklinks=true}
\providecommand{\doi}[1]{%
  \href{#1}{\nolinkurl{#1}}%
}

\def\tsc#1{\csdef{#1}{\textsc{\lowercase{#1}}\xspace}}
\tsc{WGM}
\tsc{QE}

\begin{document}
\let\WriteBookmarks\relax
\def\floatpagepagefraction{1}
\def\textpagefraction{.001}

\shorttitle{}    

\shortauthors{}  

\title [mode = title]{Anisotropic spin splitting in \texorpdfstring{$\alpha$}{alpha}-MnTe epilayers on InP(111)A and Si(111) substrates}  


\author[1]{Zofia Kopacz-Maj}[orcid = 0000-0001-7767-8555]
\credit{Conceptualisation, Investigation, Data curation, Visualisation, Writing - original draft}

\author[1]{Ali Akbar}[orcid = 0009-0008-2794-2491]
\credit{Investigation, Visualisation, Writing - original draft}

\author[1]{Aleksandr Kazakov}[orcid = 0000-0002-8066-2922]
\credit{Investigation, Conceptualisation, Writing - review \& editing}

\author[1]{Bart\l{}omiej Turowski}[orcid = 0000-0003-1229-4416]
\credit{Investigation}

\author[1,2]{Wojciech Zaleszczyk}[orcid = 0000-0002-0051-7267]
\credit{Investigation}

\author[2]{Marta Aleszkiewicz}[orcid = 0000-0002-4939-9855]
\credit{Investigation}

\author[1]{Tomasz Wojciechowski}[orcid = 0000-0002-6424-988X]
\credit{Methodology}

\author[3]{Rafa\l{} Kurleto}[orcid =  0000-0002-4749-1084]
\credit{Methodology}

\author[3]{Dawid Wutke}[orcid = 0000-0003-4355-5025]
\credit{Methodology, Software}

\author[3]{Natalia Olszowska}[orcid = 0000-0001-8377-4615]
\credit{Methodology}

\author[1]{Tomasz Wojtowicz}[orcid = 0000-0003-4498-4622]
\credit{Resources, Project administration}

\author[1,4]{Valentine V. Volobuev}[orcid = 0000-0002-8474-7392]
\ead{Volobuiev@MagTop.ifpan.edu.pl}
\cormark[1]
\credit{Investigation, Conceptualisation, Writing - review \& editing, Supervision}

\affiliation[1]{organization={International Research Centre MagTop, Institute of Physics, Polish Academy of Sciences},
            addressline={Al. Lotników 32/46}, 
            city={Warsaw},
          citysep={}, 
            postcode={02-668}, 
            country={Poland}}

\affiliation[2]{organization={Institute of Physics, Polish Academy of Sciences},
            addressline={Al. Lotników 32/46}, 
            city={Warsaw},
          citysep={}, 
            postcode={02-668}, 
            country={Poland}}

\affiliation[3]{organization={National Synchrotron Radiation Centre SOLARIS, Jagiellonian University},
            addressline={Czerwone Maki 98}, 
            city={Cracow},          
            citysep={}, 
            postcode={30-392}, 
            country={Poland}}

\affiliation[4]{organization={National Technical University “KhPI”},
            addressline={Kyrpychova Str. 2}, 
            city={Kharkiv},          
            citysep={}, 
            postcode={61002}, 
            country={Ukraine}}

\cortext[1]{Corresponding author}


\nonumnote{}

\begin{abstract}
$\alpha$-MnTe has emerged as a prototypical altermagnetic material, combining compensated antiferromagnetic order with momentum-dependent spin splitting. Here, we investigate the structural and electronic properties of epitaxial $\alpha$-MnTe layers grown by molecular beam epitaxy on InP(111)A and Si(111) substrates. The crystalline and surface quality of the epilayers was characterised by reflection high-energy electron diffraction (RHEED), atomic force microscopy (AFM), and X-ray diffraction (XRD), confirming the formation of high-quality, single-phase films. The transport measurements were performed, confirming an antiferromagnetic transition and weak ferromagnetism. The electronic structure was studied using angle-resolved photoemission spectroscopy (ARPES). At 90~K, both samples exhibit relativistic anisotropic valence band splitting, which is substantially larger along the $\Gamma$-K direction, reaching approximately 300~meV, than along $\Gamma$-M, where values of about 150~meV are observed. The magnitude and momentum dependence of the splitting are very similar for $\alpha$-MnTe grown on the two different substrates, indicating the robustness of the altermagnetic electronic structure with respect to the epitaxial environment.  
\end{abstract}

\begin{highlights}
\item High quality growth of $\alpha$-MnTe on InP(111)A and Si(111)substrates.
\item ARPES directly compares altermagnetic splitting in $\alpha$-MnTe grown on InP(111)A and Si(111).
\item The splitting disappears at 300~K, linking it to the magnetically ordered phase.
\item The results demonstrate robustness of the $\alpha$-MnTe electronic structure against the epitaxial environment once the epilayers are relaxed.
\end{highlights}


\maketitle

\section{Introduction}\label{introduction}
Altermagnets are a recently established class of magnetic materials, gaining significant attention within condensed matter physics. They combine a compensated magnetic order and momentum-dependent spin splitting of the electronic bands. In altermagnets, collinear opposite spin sublattices are connected by rotational or improper rotational symmetry, resulting in spin-split electronic bands even without spin-orbit coupling, while exhibiting a vanishing net magnetisation. The spin-splitting alternates across momentum space and vanishes along symmetry-protected nodal planes. \cite{Hayami2019, Mazin2022, Turek2022, Yuan2020, Smejkal2022_1, Smejkal2022_2, Krempasky2024}

A combination of spin-split electronic bands, as in ferromagnets, with vanishing net magnetisation, as in antiferromagnets, makes altermagnets attractive for spintronic applications. The compensated magnetic order suppresses stray fields and magnetic cross-talk, while allowing the generation and manipulation of spin currents. Other research areas employing altermagnets are i.a. valleytronics, spin caloritronics and magnonics. \cite{Smejkal2022_2, Jungwirth2026, Gonzalez-Hernandez2021, Shen2025, Liang2026, Zhou2024, Yi2025, Smejkal2023, Liu2024}

Hexagonal $\alpha$-MnTe is a model altermagnetic semiconductor. It has the NiAs-type crystal structure with P6\textsubscript{3}/mmc space group \cite{Kriegner2016, Lee2025, Takahashi2025}. It is intrinsically p-type, mainly due to native acceptor defects, particularly Mn vacancies, placing the Fermi level near or within the valence band \cite{Lee2025, Zhang2026}. Below its Néel temperature of around 310 K, the Mn moments align parallel within the basal plane along <1$\overline{1}$00> easy axis direction and antiparallel between adjacent planes, resulting in vanishing magnetisation \cite{Kriegner2016, Kriegner2017, Amin2024}. The opposite spin sublattices in $\alpha$-MnTe are connected by a nonsymmorphic sixfold screw operation \{C\textsubscript{6z}\textbar001/2\} \cite{Amin2024, Lee2025, Takahashi2025}. In addition to its altermagnetic electronic structure, $\alpha$-MnTe exhibits a weak ferromagnetic component associated with relativistic spin-orbit coupling. In particular, a staggered Dzyaloshinskii–Moriya interaction can produce a small canting of the Mn moments, resulting in a finite net magnetization and contributing to the field-dependent magnetic and transport properties of MnTe \cite{Kluczyk2024, mazin2024origin, autieri2025staggered, ChenYe2026, liu2026observation}.

Beyond its fundamental interest, $\alpha$-MnTe is also a promising material for spintronic applications. Its high Néel temperature, large momentum-dependent spin splitting, and compensated magnetic order provide a combination of properties that is attractive for spin-current generation and manipulation. In addition, the anomalous Hall effect observed in epitaxial $\alpha$-MnTe films demonstrates the possibility of exploiting altermagnetic electronic structure in electrical transport. \cite{Jungwirth2026, Krempasky2024, Amin2024, zhou2026surface, hirakida2026multipole} These functionalities are particularly relevant in thin-film geometries, where epitaxial strain, interfaces and reduced dimensionality can modify the magnetic and electronic properties. \cite{Dittmar2026, Lee2025} For eventual device integration, it is therefore important to achieve high-quality $\alpha$-MnTe films on technologically relevant substrates. In particular, Si(111) is an attractive platform because of its central role in semiconductor technology, and recent work has demonstrated epitaxial $\alpha$-MnTe on Si(111) and highlighted its potential as a silicon-compatible altermagnetic system. \cite{Lee2025, sarkar2026anomalous}

In the present work, we investigate the growth, structural, transport, and electronic properties of $\alpha$-MnTe films grown by molecular beam epitaxy (MBE) on InP(111)A and Si(111) substrates. Growth of $\alpha$-MnTe on Si(111) is particularly challenging because of the large lattice mismatch between the two materials. Previous studies on Si have been either focused on ultrathin MnTe layers \cite{Lee2025} or reported the coexistence of different structural phases \cite{sarkar2026anomalous}. Here, we demonstrate the growth of relatively thick, pure phase of NiAs-type $\alpha$-MnTe films on Si(111). In addition, the MnTe/Si(111) films exhibit hysteretic magnetotransport behaviour, most likely associated with the weak ferromagnetic component of $\alpha$-MnTe.
The electronic structure is studied using angle-resolved photoemission spectroscopy (ARPES), focusing on momentum-dependent band splitting. A comparison of $\alpha$-MnTe films grown on the two substrates allows us to examine the influence of the epitaxial environment on the altermagnetic electronic structure and to determine whether the magnitude and momentum dependence of the spin splitting are preserved. These results provide insight into the substrate dependence of the electronic properties of epitaxial $\alpha$-MnTe and the suitability of different substrate platforms for its integration in thin-film structures.

\section{Experimental details}\label{experiment}
At first, the growth conditions for epitaxial $\alpha$-MnTe were optimized for InP(111)A and Si(111) substrates in order to obtain smooth, single-phase films with high crystalline quality. The optimized growth parameters were subsequently used to prepare the samples investigated in the present study.

The MnTe films were grown by molecular beam epitaxy (GENxplor MBE system from VEECO) under an ultra-high vacuum. Prior to growth, both substrates were glued to tantalum plates using an eutectic Ga-In mixture, welded by Ta stripes and pre-baked at 150\textsuperscript{$\circ$}C in load-lock. High-purity 5N7 Mn and 6N Te were simultaneously evaporated from elemental sources. Epitaxial $\alpha$-MnTe films with a thickness of 300~nm on InP(111)A and 367~nm on Si(111) were successfully grown at a substrate temperature of 370~\textsuperscript{$\circ$}C, as measured by a calibrated pyrometer. The growth was monitored in situ by reflection high-energy electron diffraction (RHEED). 

Following the growth, the samples were transferred directly to the ARPES chamber under ultra-high vacuum conditions in the UHV suitcase, without breaking vacuum, thereby avoiding exposure to ambient conditions. ARPES measurements were performed at the URANOS beamline of the SOLARIS National Synchrotron Radiation Centre equipped with a Scienta DA30L electron energy analyser, which has an energy resolution of 1.8 meV and an angular resolution of 0.1\textsuperscript{$\circ$} \cite{Olszowska2026}. For core level measurements, a photon energy of 120~eV was used. The ARPES maps were acquired using 21~eV photons, providing enhanced surface sensitivity and probing a $k_z$ value close to the $\Gamma$ point \cite{Hajlaoui2024}. Linear horizontal polarisation was employed. Measurements were performed at 90~K and 300~K.

After completion of the ARPES measurements, the samples were removed from the vacuum system and subjected to ex situ structural and transport characterization. The crystalline structure and surface morphology were examined by X-ray diffraction (XRD) and atomic force microscopy (AFM). Electrical transport measurements were performed in a Quantum Design Physical Property Measurement System (PPMS) using a Keithley delta-mode measurement configuration with an excitation current of 1~$\mu$A. Hall bar geometry defined along [112] direction of the substrate using optical lithography was used. The measurements were carried out over the 2-250~K temperature range and in magnetic fields up to $\pm$9~T, perpendicular to the film's surface. The structural and transport properties of films grown under optimized conditions are discussed in the following sections.

Our experimental sequence ensured that the pristine as-grown surface was investigated by ARPES before any exposure to ambient conditions or ex situ processing.

\section{Growth and structural characterization}\label{growth_characterisation}
\begin{figure}
    \centering
    \includegraphics[width=0.8\linewidth]{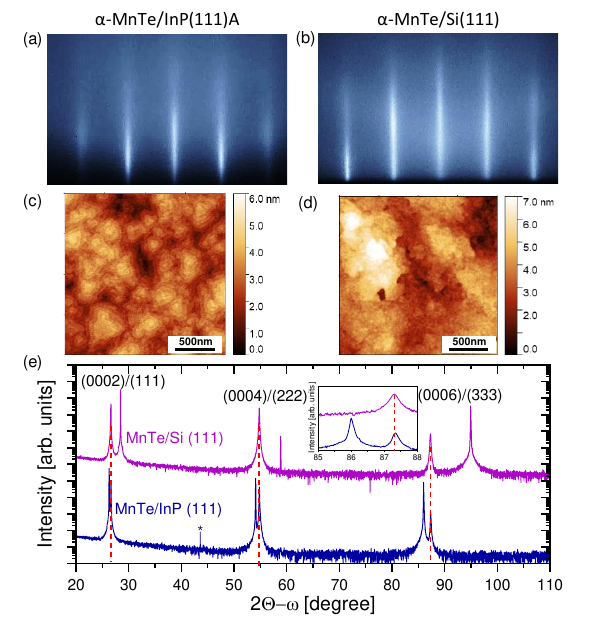}
    \caption{Structural characterisation of the samples. In situ reflection high energy electron diffraction (RHEED) images of  300~nm and 367~nm thick epitaxial $\alpha$-MnTe films grown on InP(111)A (a),  and Si(111) (b) taken along [11$\overline{2}$0] of $\alpha$-MnTe. (c, d) Smooth surfaces of MnTe/InP and MnTe/Si films respectively shown by atomic force microscopy scan. (e) X-ray diffraction (2$\Theta$-$\omega$) scan of the films shows (0001) orientation on both used substrates. The peaks corresponding to $\alpha$-MnTe are marked with red dashed line. The Bragg peak around (0006) of  MnTe, magnified in the inset, correspond to  nearly identical lattice constant of $\alpha$-MnTe films on two different substrates. An additional peak in MnTe/InP scan marked with an asterisk is due to Renninger effect.}
    \label{fig:gr_str}
    
\end{figure}
The optimized growth conditions yield smooth, two-dimensional MnTe epilayers, as evidenced by the streaky RHEED patterns shown in Fig. \ref{fig:gr_str} (a, b). The surface morphology examined by AFM is presented in Fig. \ref{fig:gr_str} (c, d). The MnTe film grown on InP(111)A exhibits a smooth, terraced surface with an RMS roughness of 0.7~nm, consistent with favourable lattice matching between $\alpha$-MnTe and InP(111), for which the nominal in-plane lattice mismatch is <1\% \cite{Kriegner2016}. Despite the much larger nominal lattice mismatch of approximately 8\% for MnTe/Si(111) \cite{Nakanishi1981}, the film on Si(1111) substrate also has a relatively smooth surface, with an RMS roughness of 1.2~nm.

The diffraction patterns from XRD study are shown in Fig. \ref{fig:gr_str}(e). The long-range 2$\Theta$-$\omega$ scan confirms growth along \textit{c} axis, which is normal to the triangular (111) surfaces of the cubic substrates. This yields the epitaxial relationship (0001)\textsubscript{MnTe}||(111)\textsubscript{InP/Si}. The structure of the layers is of NiAs-type. 
No extra peaks corresponding to residual or secondary phases of MnTe are seen, confirming the crystalline quality of the prepared films. The out-of-plane lattice parameters of the MnTe films calculated from the (0006) Bragg peaks are c~=~6.694~Å for MnTe/InP and c~=~6.696~Å for MnTe/Si, as shown in the inset of Fig. \ref{fig:gr_str} (e). These values are approximately 0.02~Å smaller than the c = 6.71~Å reported for relaxed MnTe films grown on SrF$_2$ (111), a substrate that provides close lattice and thermal-expansion matching to MnTe and for which the films were reported to be nearly fully relaxed \cite{Kriegner2016, Kriegner2017}. The slightly reduced c lattice parameter observed here may therefore reflect residual strain associated with the substantially large thermal-expansion mismatch between MnTe and the InP and Si substrates. In fact, InP and Si have thermal expansion coefficients of $0.5 \times 10^{-5}$ K\textsuperscript{-1} \cite{glazov2001thermal} and $2.6 \times 10^{-6}$ K\textsuperscript{-1} \cite{Yasumasa1984} respectively, both substantially lower than that of MnTe, ($1.62 \times 10^{-5}$ K\textsuperscript{-1}) \cite{Minikayev2015}. Thus, upon cooling from the growth temperature, MnTe tends to contract more strongly than either substrate, which can generate tensile biaxial strain in the MnTe film plane and, through elastic coupling, a corresponding reduction of the out-of-plane c lattice parameter. The difference of only 0.002 Å between the MnTe/InP and MnTe/Si lattice parameters is, however, very small and approaches the experimental uncertainty. This suggests that, despite the substantially different lattice mismatches with the substrates, the lattice parameters of the relatively thick MnTe films are very close and influenced more strongly by thermal-expansion mismatch than by lattice mismatch.


\section{Transport}\label{transport}
\begin{figure}
    \centering
    \includegraphics[width=0.75\linewidth]{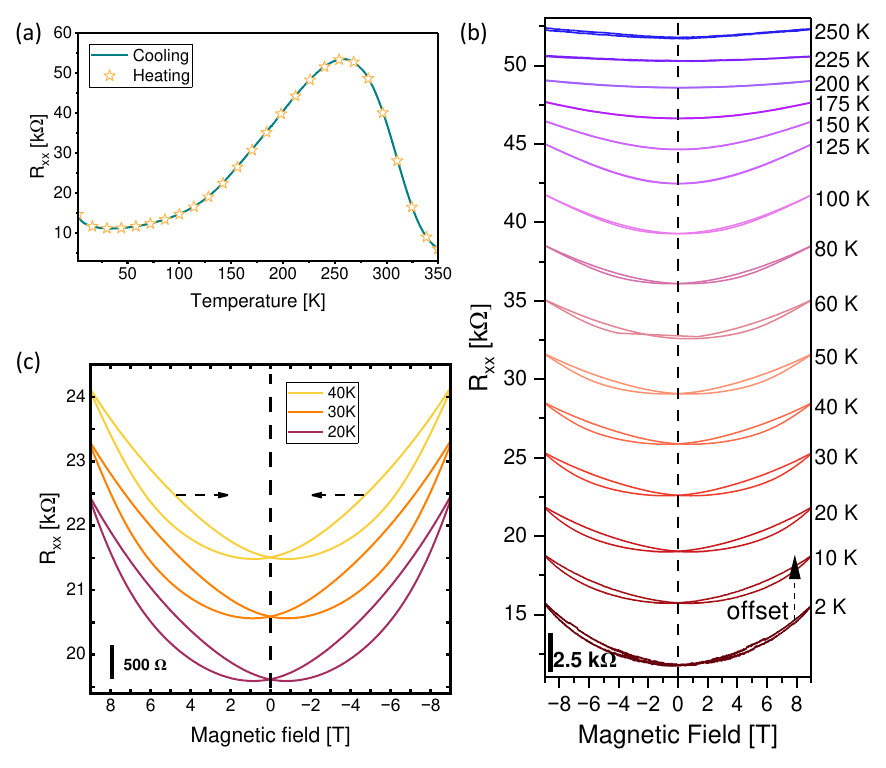}
    \caption{Electric transport properties of $\alpha$-MnTe on Si(111) substrate. (a) Zero field temperature dependence of longitudinal resistance $R_{xx}$ upon cooling (solid line) and heating (stars). (b) Magnetic field dependence for the field applied perpendiculary to the film's surface for temperatures 2~-~250~K, vertically offset for clarity. (c) Low-temperature traces of field dependence, same as in (b), in which maximum hysterises at 30K can be seen. Dashed arrows show magnetic field scan direction.}
    \label{fig:transport}
\end{figure}

The magnetotransport measurements were performed to probe the electrical and magnetic response of $\alpha$-MnTe/Si. Fig. \ref{fig:transport} (a) shows the zero field longitudinal resistance $R_{xx}$ as a function of temperature for both cooling and heating, where the dependencies coincide. The resistance exhibits a broad maximum around 250~-~260~K. The maximum in $R_{xx}(T)$ is consistent with a change in the dominant scattering mechanism near the antiferromagnetic phase transition. Similar temperature dependences have been attributed to enhanced spin disorder and critical spin fluctuation scattering close to the Néel temperature \cite{Wasscher1965, Magnin2012, Kriegner2016}. At temperatures below approximately 50 K, the resistance exhibits an upturn upon cooling, which can be attributed to low-temperature carrier localization and freeze-out in the semiconducting MnTe films. Such behaviour has been reported for $\alpha$-MnTe and is generally associated with impurity- or defect-assisted transport and the crossover from extended-band conduction to localized/hopping conduction at low temperatures. \cite{Woochul2009, Gao2026}

Fig. \ref{fig:transport} (b) shows the magnetoresistance results for temperatures between 2 and 250~K, with curves vertically offset for clarity. At low temperatures, a clear difference between the increasing and decreasing field sweeps is observed, resulting in a hysteretic magnetoresistance response. The hysteresis weakens upon warming up. The electronic response of the film is consistent with that expected for magnetically ordered $\alpha$-MnTe. Low-temperature hysteresis indicates weak ferromagnetism, which has been reported previously for $\alpha$-MnTe \cite{Wasscher1965, Chilcote2024, Kluczyk2024} and assigned to spin canting due to spin-orbit coupling \cite{ChenYe2026}. 

The transport measurements complement the structural results by confirming that the film exhibits the characteristic magnetic and electronic response of $\alpha$-MnTe, while the low-temperature hysteresis points to a weak ferromagnetism superimposed on the dominant antiferromagnetic state. 

\section{ARPES}\label{ARPES}
\subsection{Core levels}\label{CLs}
\begin{figure}
    \centering
    \includegraphics[width=0.7\linewidth]{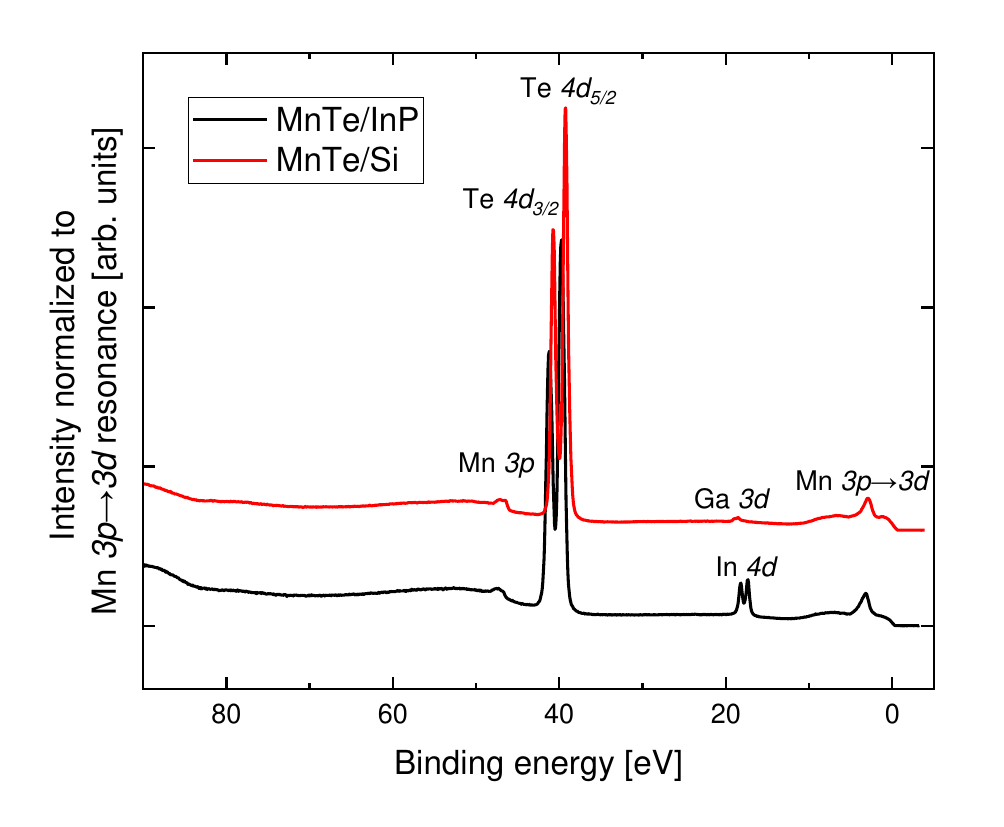}
    \caption{Core levels of $\alpha$-MnTe on InP(111)A substrate (black) and Si(111) (red) with signed photoemission features. Intensity is normalised to the intensity of Mn \textit{3p}$\rightarrow{}$\textit{3d} resonance feature and offset for clarity.}
    \label{fig:CLs}
\end{figure}

Fig. \ref{fig:CLs} compares the core level photoemission spectra of the $\alpha$-MnTe films grown on InP(111)A and Si(111), measured using a photon energy of 120~eV. The spectra are normalised to the Mn \textit{3p}$\rightarrow{}$\textit{3d} resonance feature around the binding energy of 3-4~eV \cite{Bressler1994, Kowalski2001} to account for the photon beam intensity shift between the measurements. The spectra are dominated by Mn- and Te-derived features. In both samples, an intense Te \textit{4d} spin-orbit doublet is observed at binding energies of approximately 39–41~eV. A broader Mn \textit{3p} feature is present at binding energies of around 45–48~eV. The observation of Te \textit{4d} and Mn \textit{3p} core level features is consistent with other studies \cite{Lee2025, Osumi2024, Ueda1994}. The similarity of the Te \textit{4d} and Mn \textit{3p} features observed for $\alpha$-MnTe grown on InP(111)A and Si(111) suggests that the dominant chemical environment of Mn and Te is comparable in the two films.

Differences between the two spectra are observed in the 17-20~eV range of binding energy. For the MnTe/InP sample, a pronounced feature at approximately 17–18~eV is assigned to the In \textit{4d} core level \cite{Schmidt1998}. Since the photoemission at h$\nu$~=~120~eV is strongly surface sensitive, the observation of a sizeable In signal through the MnTe layer suggests In diffusion from the InP substrate. A contribution from the eutectic Ga-In mixture used to glue the sample to the tantalum plate cannot be completely excluded. However, the absence of a comparably strong In \textit{4d} feature for the MnTe/Si sample suggests that the InP substrate is the dominant source of the pronounced In signal in the MnTe/InP core levels spectrum. 
For MnTe/Si, only a much weaker feature is observed at approximately 18.5-19~eV. This energy is close to the Ga \textit{3d} core level \cite{Muller2026}. The weak feature may therefore originate from trace Ga contamination associated with the eutectic Ga–In paste. The absence of a clearly resolved In-derived feature in the MnTe/Si sample does not exclude this interpretation, since eutectic Ga–In is Ga rich and its surface composition is strongly dominated by Ga-containing species \cite{Dickey2008}.

The small rigid shift of the Mn-, Te-and Mn resonance-related spectral features between MnTe/InP and MnTe/Si may reflect a slight difference in the $E_{\mathrm{F}}$ associated with different doping levels in the two samples, possibly coming from an unintentional In doping. 

\subsection{Temperature dependence of electronic structure}\label{ARPES_T_dep}
\begin{figure}
    \centering
    \includegraphics[width=0.75\linewidth]{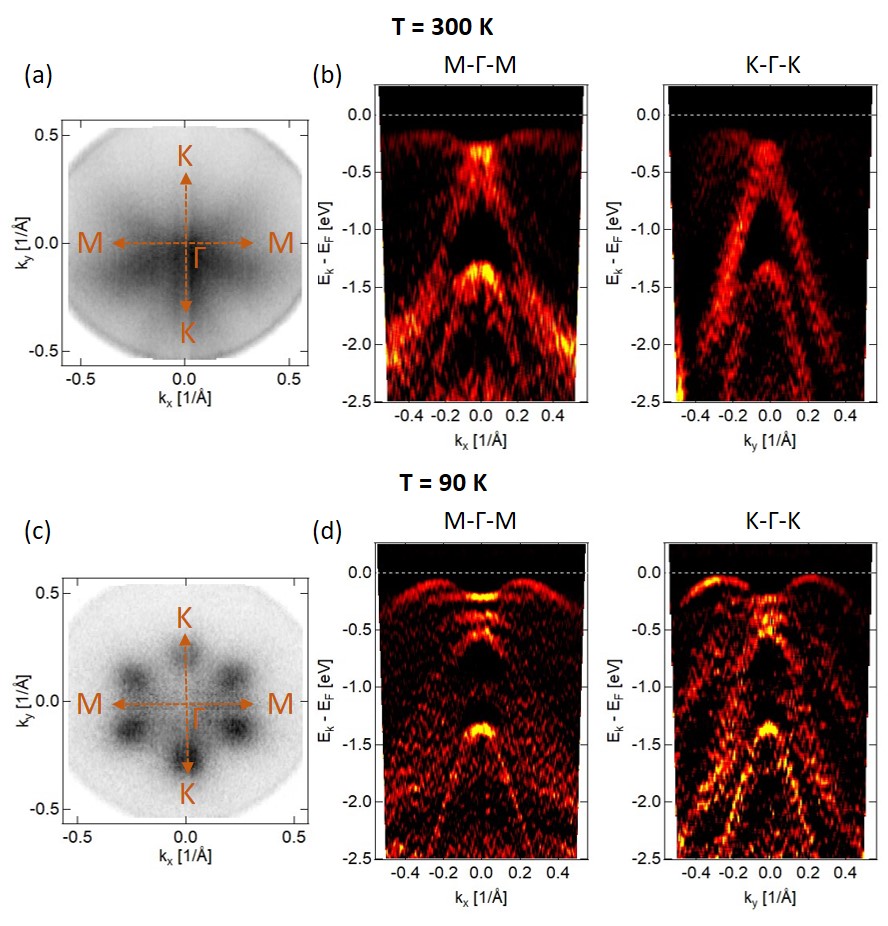}
    \caption{Temperature dependent ARPES spectra of $\alpha$-MnTe on InP(111)A at h$\nu$~=~21~eV photon energy. (a) Fermi surface and (b) curvature-enhanced spectra along M-$\Gamma$-M (left) and K-$\Gamma$-K (right) directions at T~=~300~K.(c) Snowflake-shaped Fermi surface map and (d) curvature-enhanced spectra along M-$\Gamma$-M (left) and K-$\Gamma$-K (right) directions at T~=~90~K.}
    \label{fig:InP_300K_90K}
\end{figure}

Fig. \ref{fig:InP_300K_90K} compares the Fermi surface maps and the ARPES dispersion of $\alpha$-MnTe on the InP(111) substrate measured at 300~K and 90~K using photon energy of 21~eV. 
For an improved visualisation of  bands in Fig. \ref{fig:InP_300K_90K} (b, d), the data were processed using the curvature method introduced by Zhang et al. \cite{Zhang2011}. This method enhances spectral weight of bands while preserving their dispersion, facilitating the identification of the underlying electronic structure.

At 300~K (Fig. \ref{fig:InP_300K_90K} (a, b)), which is close to the Néel temperature, the Fermi surface map exhibits a comparatively broad spectral features with the underlying hexagonal symmetry visible. The corresponding {energy-momentum cuts along M-$\Gamma$-M and K-$\Gamma$-K directions reveal broad valence band dispersions, and the states close to $E_{\mathrm{F}}$ are weakly resolved. After cooling to 90~K (Fig. \ref{fig:InP_300K_90K} (c, d)) the Fermi surface} map shows snowflake-shaped momentum dependence. The valence band statees sharpen. Most importantly, the bands near $E_{\mathrm{F}}$ evolve into multiple branches, revealing a band splitting along M-$\Gamma$-M and K-$\Gamma$-K directions. 

The emergence of the band splitting at low temperature is consistent with previous temperature-dependent ARPES studies of $\alpha$-MnTe and supports the association of band splitting with the strengthening of the magnetically ordered altermagnetic state. Considering that the measurements probe $k_z$ values close to zero, together with the onset of splitting below the magnetic ordering temperature, indicates a relativistic altermagnetic origin of the observed band splitting \cite{Krempasky2024, Hajlaoui2024}. Hajlaoui et al. performed temperature-dependent ARPES at the same photon energy, h$\nu$~=~21~eV and observed a progressive enhancement of the valence band splitting on cooling below the Néel temperature. In that study, the splitting reached 150~meV along $\Gamma$-K direction at 50~K and became unresolved above approximately 270~K. At low temperature, the band splitting observed in our data shows a close resemblance to that reported by Hajlaoui et al. In their study, the corresponding split valence-band features were assigned to bulk electronic states \cite{Hajlaoui2024}. A closely related temperature-dependent lifting of band degeneracy was also reported by Lee et al., who observed the development of a momentum-dependent splitting below the magnetic transition and reported splitting reaching 370~meV along D-U direction, defined as a direction parallel to $\Gamma$-M at $k_z$ = 0,3$\pi$/c \cite{Lee2024}.

Together with the structural identification of the NiAs-type $\alpha$-MnTe phase and the characteristic Mn and Te core level spectra discussed above, the ARPES results provide complementary electronic structure evidence for the formation of $\alpha$-MnTe and demonstrate that its characteristic low-temperature altermagnetic band structure is preserved.

\subsection{Substrate dependence of the low-temperature electronic structure}\label{ARPES_low_T}
\begin{figure}
    \centering
    \includegraphics[width=0.75\linewidth]{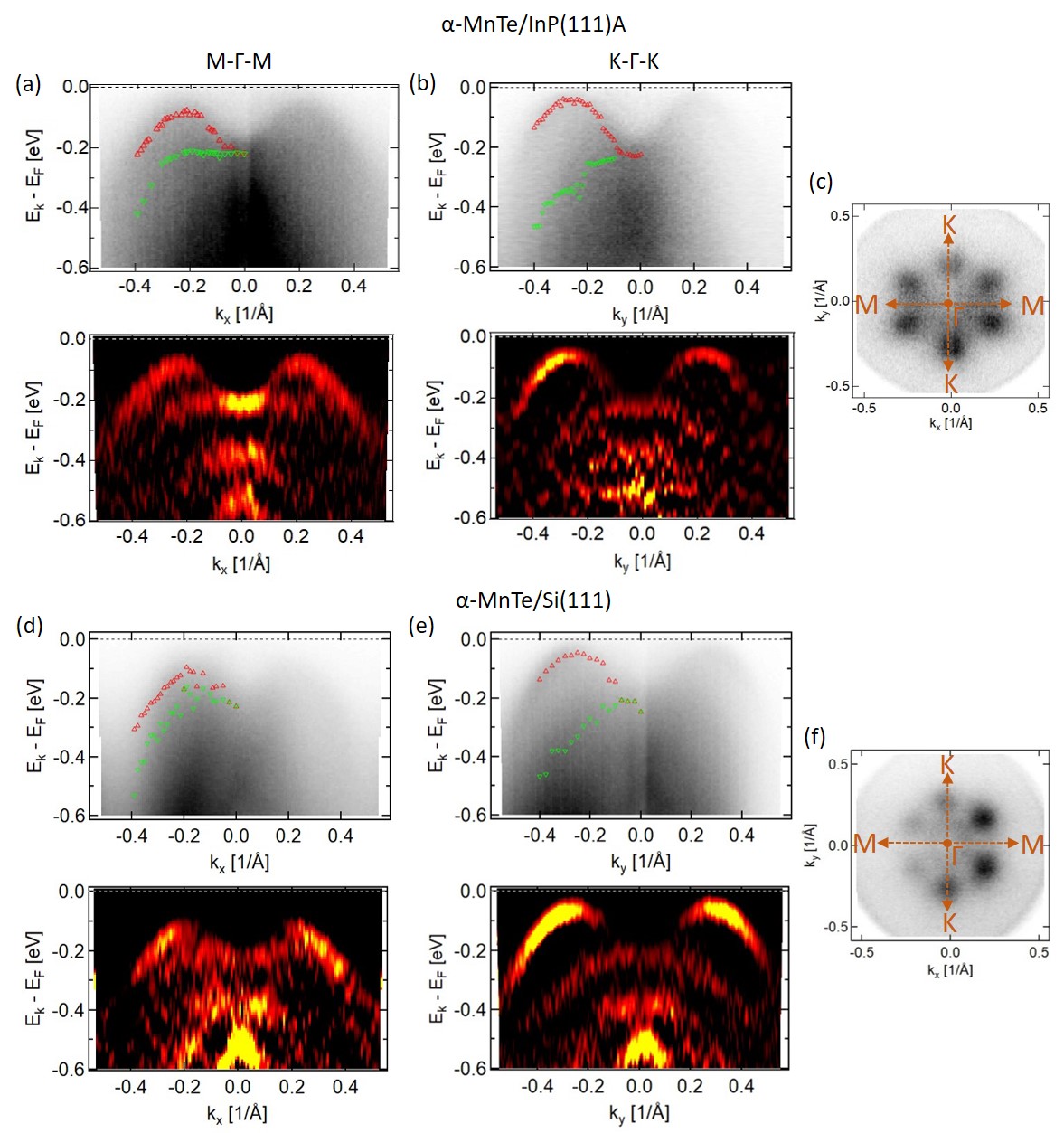}
    \caption{ARPES spectra of $\alpha$-MnTe measured at h$\nu$~=~21~eV, T~=~90~K. (a, d) Energy momentum cuts in M-$\Gamma$-M direction and (b, e) along in K-$\Gamma$-K direction for $\alpha$-MnTe grown on InP(111)A and Si(111), respectively. For each data set, the upper panels show the raw ARPES intensity, with red and green markers indicating the positions of split bands fitted with Lorentzian peaks in EDCs, while the lower panels show the corresponding curvature-enhanced representations. (c,f) Constant-energy maps for the InP(111)A and Si(111) samples, respectively, with the M--$\Gamma$--M and K--$\Gamma$--K directions indicated by orange lines.}
    \label{fig:InP_Si}
\end{figure}
Having demonstrated that the band splitting develops upon cooling into the magnetically ordered state, we examine whether the momentum dependence is preserved when $\alpha$-MnTe is grown on different substrates. Fig. \ref{fig:InP_Si} summarises the ARPES data along M-$\Gamma$-M and K-$\Gamma$-K high-symmetry directions at 90~K. The upper panels in Fig. \ref{fig:InP_Si} (a, b, d, e) show the raw ARPES intensity distributions, while the lower panels display the same data after curvature enhancement of the bands. In addition, to better visualise the band splitting, energy dispersion curves (EDCs) profiles were extracted from the raw ARPES data and analysed. The EDCs contain a substantial smoothly varying background that masks the comparatively weak MnTe valence band features near the $E_{\mathrm{F}}$. In the range of 0~-~0.6~eV below $E_{\mathrm{F}}$ the slowly varying background was modelled as a third order polynomial and the background was removed. Lorentzian curves were then fitted to the split band maxima of EDCs and their positions were superimposed on the raw data as red and green markers. The curvature maps below the raw spectra reproduce the fitted band dispersions.

For the $\alpha$-MnTe/InP(111)A sample (Fig. \ref{fig:InP_Si} (a, b)), two distinct branches near $E_{\mathrm{F}}$ are resolved along both crystallographic directions. The separation between the fitted peak positions increases away from the $\Gamma$ point, demonstrating the momentum-dependent character of the splitting. A closely similar behaviour is observed for the $\alpha$-MnTe/Si(111) sample. Despite the different substrate, the Lorentzian derived peak positions reveal similar dispersions. Calculations by Krempaský et al. established that the split bands possess opposite spin polarization, linking this band separation to the relativistic lifting of Kramers spin degeneracy in $\alpha$-MnTe \cite{Krempasky2024}.

The Fermi surfaces on both substrates (Fig. \ref{fig:InP_Si} (c, f)) reflect the underlying hexagonal symmetry. They exhibit a characteristic snowflake-like contour centred at $\Gamma$. The deviation from a circular contour reflects the in-plane anisotropy of the valence band electronic structure. While the underlying $\alpha$-MnTe crystal structure has hexagonal symmetry, the electronic structure of a single magnetic domain is expected to exhibit reduced, approximately twofold symmetry near the valence-band maximum due to the in-plane orientation of the Néel vector \cite{Krempasky2024}. The predominantly sixfold-symmetric contours observed here may therefore indicate that the ARPES measurements probe multiple magnetic domains \cite{Amin2024, Kriegner2017}, whose contributions restore the apparent sixfold symmetry.

\begin{figure}
    \centering
    \includegraphics[width=0.65\linewidth]{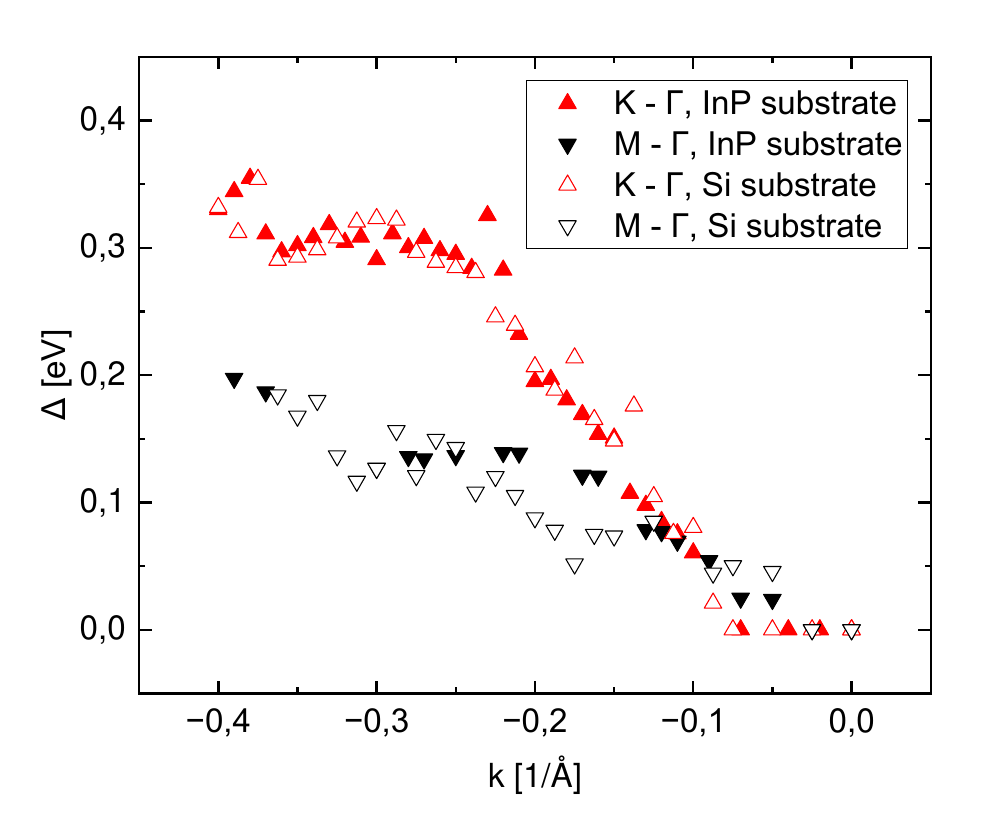}
    \caption{Band splitting at h$\nu$~=~21~eV, T~=~90~K. The splitting is taken as the difference of the positions of the Lorentz curves fitted to the unprocessed ARPES maps in K-$\Gamma$, M-$\Gamma$ directions for samples on both InP(111)A and Si(111) substrates.}
    \label{fig:splitting}
\end{figure}

To quantify the momentum-dependent band separation observed in Fig. \ref{fig:InP_Si}, the energy splitting $\Delta E$ was determined from the difference between the peak positions obtained by Lorentzian fitting of the EDCs extracted from raw data. The resulting $\Delta E(k)$ values are summarised in Fig. \ref{fig:splitting} for both K-$\Gamma$ and M-$\Gamma$ directions and both InP(111)A and Si(111) substrates. For both substrates, the splitting decreases continuously towards the $\Gamma$ point and exhibits pronounced in-plane anisotropy away from $\Gamma$. Along K–$\Gamma$, the splitting reaches approximately 300~meV, comparable to the value reported by Osumi et al. \cite{Osumi2024}, whereas substantially smaller values of approximately 150~meV are obtained along M–$\Gamma$. 

The pronounced in-plane anisotropy observed in Fig. \ref{fig:InP_Si},\ref{fig:splitting} is consistent with previous ARPES studies of $\alpha$-MnTe near $k_z$ = 0 nodal plane, which reported a substantially larger splitting along $\Gamma$-K than along $\Gamma$-M  \cite{Krempasky2024, Hajlaoui2024, Zhang2026}. However, the magnitude of the splitting observed in the present work is substantially larger than that reported in several of these studies. In particular, the $\Gamma$–K splitting of approximately 300~meV obtained here is about a factor of two larger than the $\sim$150~meV reported by Hajlaoui et al. under nominally similar photon-energy conditions \cite{Hajlaoui2024}. The origin of the quantitative difference in the splitting magnitude between experiments is therefore of particular interest. One possible reason can be the difference in magnetic-domain configuration probed by ARPES. Since the electronic structure and consequently the band splitting of $\alpha$-MnTe depends on the orientation of the Néel vector \cite{Krempasky2024, Hajlaoui2024}, the measured spectrum from a multidomain sample represents a superposition of contributions from domains with different magnetic orientations. Differences in domain populations and their relative spectral weights can therefore modify the apparent band separation along a particular momentum direction. 

Importantly, however, within the present study the magnitude and anisotropy of the splitting are very similar for MnTe grown on InP(111)A and Si(111), despite the substantially different epitaxial environments. This indicates that the observed momentum dependence is robust with respect to the substrate choice in the relaxed films. Therefore, the present measurements provide a direct comparison showing that the characteristic anisotropic splitting is retained in relaxed $\alpha$-MnTe films grown in substantially different epitaxial environments.

\section{Conclusions}
In conclusion, our results demonstrate that the characteristic anisotropic band splitting of $\alpha$-MnTe is preserved in relaxed epilayers grown on both InP(111)A and Si(111), despite their substantially different epitaxial environments. The close similarity of the low-temperature ARPES spectra collected on the two types of epilayers shows that the altermagnetic electronic structure is robust against substrate choice, while the structural and magnetotransport measurements confirm the expected $\alpha$-MnTe phase and magnetic response. This substrate robustness makes both InP(111)A and Si(111) promising platforms for the integration of altermagnetic MnTe into future electronic and spintronic device architectures.

\section{}\label{}

\printcredits

\section{Acknowledgements}
Publication subsidized from the state budget within the framework of the programme of the Minister of
Science (Polska) called Polish Metrology II Project No. PMII/
SP/0012/2024/02, amount of subsidy 944,900.00 PLN, total
value of the Project 944,900.00 PLN. This research was also partially supported by the ``MagTop'' project (FENG.02.01-IP.05-0028/23) carried out within the ``International Research Agendas'' programme of the Foundation for Polish Science co-financed by the European Union under the European Funds for Smart Economy 2021-2027 (FENG). A.A., V.V.V. and A.K. also acknowledge Narodowe Centrum Nauki (NCN, National Science Centre, Poland) IMPRESS-U Project No. 2023/05/Y/ST3/00191. V.V.V. thanks long-term program of support of the Ukrainian research teams at the Polish Academy of Sciences carried out in collaboration with the U.S.
National Academy of Sciences with the financial support of external partners.

The authors acknowledge the financial support provided by the Polish Ministry of Science and Higher Education through SPUB subsidies for financing the maintenance of the Molecular Beam Epitaxy (MBE) Laboratory of Semiconductor Nanostructures (31/623767/SPUB/SP/2025) and the Electron Microscopy and Nanolithography Laboratory (21/599012/SPUB/SP/2024).

Research at the National Synchrotron Radiation Centre SOLARIS is supported by the Ministry of Science and Higher Education, Poland, under contract no. MNiSW/2026/DWM/928.

We thank Dr. Marcin Rosmus for kindly sharing the Igor Pro scripts used to process the ARPES data. 

\bibliography{cas-refs}

\end{document}